\documentclass[12pt]{iopart}
\usepackage{graphicx}
\usepackage{iopams} 
\begin{document}

\title{Balanced Homodyne Detection of Optical Quantum States at Audio-Band Frequencies and Below}

\author{M S Stefszky,$^{1}$ C M Mow-Lowry,$^{1}$ S S Y Chua,$^{1}$ D A Shaddock,$^{1}$ B C Buchler,$^{1}$ H Vahlbruch,$^{2}$ A Khalaidovski,$^{2}$ R Schnabel,$^{2}$ P K Lam,$^{1}$ and D E McClelland$^{1}$}

\address{$^1$Department of Quantum Science, \\ The Australian National University,
ACT 0200, Australia}
\address{$^2$Institut f$\ddot{\mathrm{u}}$r Gravitationsphysik of Leibniz Universit$\ddot{\mathrm{a}}$t Hannover and Max-Planck-Institut f$\ddot{\mathrm{u}}$r Graviationsphysik (Albert-Einstein-Institut), Callinstr. 38, 30167 Hannover, Germany}

\ead{michael.stefszky@anu.edu.au}

\begin{abstract}
The advent of stable, highly squeezed states of light has generated great interest in the gravitational wave community as a means for improving the quantum-noise-limited performance of advanced interferometric detectors. To confidently measure these squeezed states, it is first necessary to measure the shot-noise across the frequency band of interest. Technical noise, such as non-stationary events, beam pointing, and parasitic interference, can corrupt shot-noise measurements at low Fourier frequencies, below tens of kilo-Hertz. In this paper we present a qualitative investigation into all of the relevant noise sources and the methods by which they can be identified and mitigated in order to achieve quantum noise limited balanced homodyne detection. Using these techniques, flat shot-noise down to Fourier frequencies below 0.5$\,$Hz is produced. This enables the direct observation of large magnitudes of squeezing across the entire audio-band, of particular interest for ground-based interferometric gravitational wave detectors. 11.6$\,$dB of shot-noise suppression is directly observed, with more than 10$\,$dB down to 10$\,$Hz.
\end{abstract}

\submitto{\CQG}

\section{Introduction}

It has long been known that squeezing could be used to improve the sensitivity of ground based gravitational wave detectors \cite{Caves81.PRD}. Only very recently, however, has this concept been demonstrated in a full scale interferometer \cite{GWOQE.NPH}. There have been a number of key advances made in the production, control and detection of squeezed states that have made this possible \cite{Vahlbruch06.PRL,McKenzie05.JOB,Schnabel10.NC}. One challenge, that is the particular focus of this paper, was the construction and characterisation of a detection system, from electronics to optical elements, that can accurately measure quantum noise suppression in the audio-frequency band.

The first measurements of squeezing in the audio-frequency band \cite{McKenzie04.PRL} were limited by excess noise at low frequencies. As the detection frequency decreased, a ``roll-up'' in the frequency spectrum appeared on both the shot-noise and squeezing measurements \cite{McKenzie04.PRL,Vahlbruch06.PRL}. This led to the interesting situation where it could not be determined whether the squeezing itself was degrading at low frequencies, or whether it was simply a limitation on the measurement device. It was later shown that it was indeed possible to measure white shot-noise at frequencies down to below one hertz \cite{Vahlbruch07.NJP}. Beam jitter, electronics, local oscillator noise coupling \cite{McKenzie07.AO}, non-stationary events in the balanced homodyne output \cite{Chua08.JPCS}, and parasitic interference \cite{Vahlbruch07.NJP,Schnabel10.NC} over the past few years have all been found to contribute to the excess noise at low frequencies. This paper details the methods by which some of these audio-frequency noise sources can be identified and methods by which all noise sources can be mitigated.

\begin{figure}[h]
  \begin{center}
  \includegraphics[width=7cm]{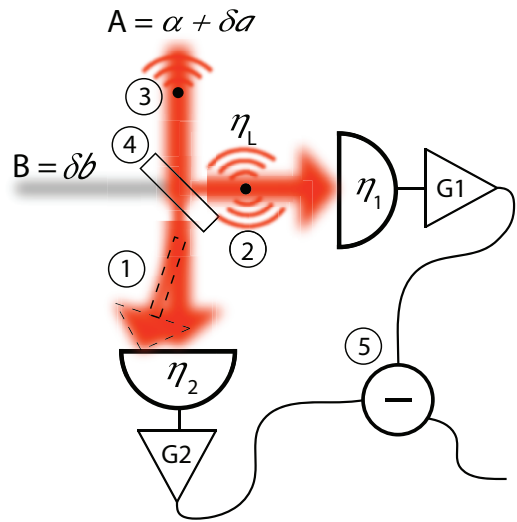}
  \end{center}
  \caption{Conceptual view of the major issues arising in low frequency measurements with a balanced homodyne setup; 1 - beam pointing,
  2 - scatter resulting in loss and/or mode shape disortion, 3 - scatter leading to parasitic interference,
  4 - the splitting ratio of the beamsplitter, 5 - the subtraction of the photocurrents. All other terms are described in the text.}  \label{Concept}
\end{figure}

After a brief introduction to balanced homodyne detection, we present a systematic investigation into all of the sources of low frequency technical noise in this detection scheme, shown in figure \ref{Concept}. We divide the issues into two groups, those of an electronic nature and those of an optical nature. Several measurements, each designed to isolate and mitigate individual sources of noise, are reported, resulting in a shot-noise measurement that is flat down to 0.5$\,$Hz. We conclude with a measurement of squeezing that is flat and shows quantum noise suppression of 10$\,$dB (which was first achieved in \cite{Vahlbruch08.PRL} at a frequency of 5$\,$MHz) or greater, down to frequencies as low as 10$\,$Hz.

\section{\bf{Noise couplings}} \label{Equations}
We illustrate the noise couplings by modelling the photocurrents in the balanced homodyne detector using the standard methods \cite{BuchlerPhD} with the inclusion of terms representing the relevant noise sources. The model is included for completeness and to provide a reference for balanced homodyne operation. The model is set up as shown in figure \ref{Concept}. A bright local oscillator and the weak signal beam of interest interfere on a beamsplitter of power splitting ratio $\eta_{{\rm bs}}$ that is close to, but not precisely, 50$\,\%$. We label the local oscillator field as $\hat{A} = \alpha + \delta \hat{a}$ and the signal field as $\hat{B} = \delta \hat{b}$, where we have assumed either a squeezed vacuum or a vacuum state, giving a coherent amplitude value of zero for the signal field. Here we will drop the hat notation for the operators. The two photo-detectors have differing quantum efficiencies, $\eta_{1}(x,y)$ and $\eta_{2}(x,y)$ that vary in the transverse plane. A loss term, $\eta_{l}$, is added to one arm of the balanced homodyne detector to simulate loss from sources such as dust passing through one of the balanced homodyne beams (location 2). We define $\delta_{V}$ terms as the vacuum fluctuation contributions entering due to the losses from the inefficient photodiodes, $\delta_{V1}$ and $\delta_{V2}$, as well as the loss in one of the balanced homodyne detector arms, $\delta_{V0}$.

The effective fields incident on the two photo-detectors, $F_{1,2}$, are written as
\begin{eqnarray}
F_{1} = \sqrt{\eta_{1}}\left(\sqrt{1-\eta_{l}}\left(\sqrt{\eta_{{\rm bs}}} A+\sqrt{1-\eta_{{\rm bs}}} B \right)+\sqrt{\eta_{l}}\delta_{V0}\right)+\sqrt{1-\eta_{1}} \delta_{V1}, \\
F_{2}=\sqrt{\eta_{2}} \left(\sqrt{\eta_{{\rm bs}}}B-\sqrt{1-\eta_{{\rm bs}}}A\right)+\sqrt{1-\eta_{2}} \delta_{V2},
\end{eqnarray}
where the explicit indication of the spatial dependence of the photodiode efficiencies have been removed for succinctness.
The photocurrent on photodiode 1 is then proportional to $F^{\dagger}_{1} F_{1}$ and similarly for the photocurrent in photodiode 2. The two photocurrents can then be written, 

\begin{eqnarray} \label{current1}
i_{{\rm HD}1} &=&\eta_{1} \eta_{{\rm bs}} (1-\eta_{l}) \alpha^{2}+\eta_{1}\eta_{{\rm bs}}(1-\eta_{l})\alpha \delta X_{a}^{+} \nonumber\\
&&+\eta_{1}\sqrt{\eta_{{\rm bs}}}\sqrt{1-\eta_{{\rm bs}}}\sqrt{1-\eta_{l}}\alpha \delta X_{b}^{+}+\sqrt{\eta_{1}}\sqrt{\eta_{{\rm bs}}}\sqrt{1-\eta_{1}}\sqrt{1-\eta_{l}}
\alpha \delta X_{V1}^{+} \nonumber\\
&&+\eta_{1}\sqrt{\eta_{{\rm bs}}}\sqrt{1-\eta_{l}}\sqrt{\eta_{l}}\alpha \delta X_{V0}^{+}
\end{eqnarray} 

\begin{eqnarray} \label{current2}
i_{{\rm HD}2} &=&\eta_{2} \sqrt{1-\eta_{{\rm bs}}} \alpha^{2}+\eta_{2}(1-\eta_{{\rm bs}})\alpha \delta X_{a}^{+}-\eta_{2}\sqrt{\eta_{{\rm bs}}}\sqrt{1-\eta_{{\rm bs}}}\alpha \delta X_{b}^{+} \nonumber\\
&&-\sqrt{\eta_{2}}\sqrt{1-\eta_{{\rm bs}}}\sqrt{1-\eta_{2}}\alpha \delta X_{V2}^{+} 
\end{eqnarray} 
where the $\delta X_{a}^{+}=\delta a^{\dagger} + \delta a$ terms are the amplitude quadrature fluctuation operators for the field fluctuation term $\delta a$ and similarly for $\delta X_{b}^{+}$. It is the subtraction of these two photocurrents, limited by experimental imperfections, that give the output of the balanced homodyne detector. Many of the coupling mechanisms for the various noise sources are described by these equations.

\section{Electronics}\label{Elec}

\subsection{\bf{Balanced Homodyne Photo-Detector Electronics Designs}}

\begin{figure}[h!]
  \begin{center}
  \includegraphics[width=8cm]{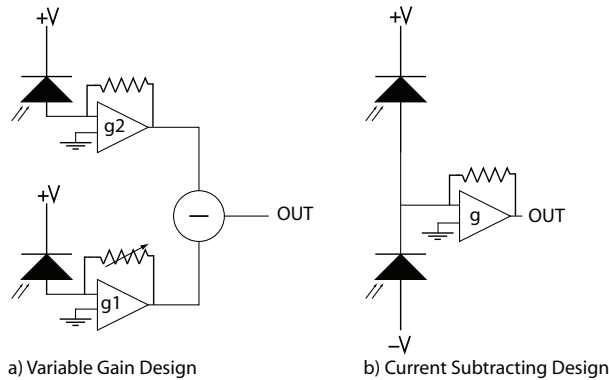}
  \end{center}
  \caption{The two balanced homodyne detector schemes: a) The variable gain design. The photocurrents from the two photodiodes have separate transimpedance gain stages, g1 and g2, and the two output signals are then subtracted. b) The current subtracting design. The two photocurrents are subtracted before undergoing any electronic gain, g.}
  \label{DetDesign}
\end{figure}

Two differing electronic designs for balanced homodyne photo-detectors were investigated. Their simplified schematics are shown in figure \ref{DetDesign}. The first is a twin photo-detector, variable gain design. In this design, the measured photocurrents are immediately amplified through a trans-impedance stage on separate electronic boards and an electronic subtractor then takes the difference between these two signals \cite{McKenzie04.PRL}. The gain between the two detectors can be varied to allow the subtraction of the photocurrents to be optimised electronically, allowing for the compensation of uneven optical powers and differences in the photodiode responses.  This allows for the compensation of unequal $\eta_{1}(x,y)$ and $\eta_{2}(x,y)$ terms as well as an imperfect $\eta_{{\rm bs}}$ term from \S \ref{Equations}. This design, to be known for the rest of this paper as the {\it variable gain} design, is given in detail in \cite{McKenziePhD}, and the first measurements of audio band squeezing utilised this style of balanced homodyne detector.

\begin{figure}[h!]
  \begin{center}
  \includegraphics[width=12cm]{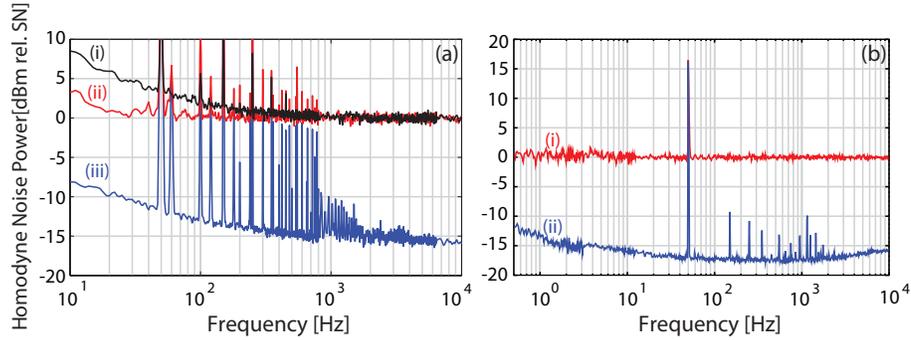}
  \end{center}
  \caption{Figure 3(a) shows the dark noise, (iii), and two shot-noise traces, (i) and (ii), for the variable gain balanced homodyne design with a local oscillator power of approximately 1.3$\,$mW. In trace (i), standard carbon film resistors are used and the shot-noise is seen to deviate strongly at lower frequencies. Replacing these resistors with low flicker noise resistors results in the near flat shot-noise measurement seen in trace (ii). The deviation seen near 10$\,$Hz is believed to be caused by scatter as it was seen to have a time dependence. Traces are pieced together from 3 FFT windows: 0-800$\,$Hz, 0-6.4$\,$kHz, 0-102.4$\,$kHz with 800 FFT lines, resulting in resolution bandwidths of 1$\,$Hz, 8$\,$Hz, and 128$\,$Hz respectively. 200 RMS averages were taken for all traces. Figure 3(b) shows the normalised dark noise (ii) and shot-noise (i) for the current subtracting design in the experimental setup described in \cite{Vahlbruch10.CQG,Khalaidovski12.ArX} with a local oscillator power of 570$\,\mu$W. Shot-noise is measured flat down to 0.5$\,$Hz. Traces are pieced together from multiple FFT windows: 0-12.5$\,$Hz, 0-50$\,$Hz, 0-200$\,$Hz, 0-400$\,$Hz, 0-800$\,$Hz, 0-1.6$\,$kHz, 0-3.2$\,$kHz, 0-6.4$\,$kHz and 0-12.8$\,$kHz, with resolution bandwidths of 62.5$\,$mHz, 250$\,$mHz, 1$\,$Hz, 2$\,$Hz, 4$\,$Hz, 4$\,$Hz, 16$\,$Hz, 32$\,$Hz and 64$\,$Hz respectively. A minimum of 100 RMS averages were taken for all traces.} \label{VolCurSN}
\end{figure}

The second design is a single electronic board design, also shown in figure \ref{DetDesign}, where the photocurrents of both photodiodes are immediately subtracted from each other before undergoing any trans-impedance amplification. In this design there is no longer a variable electronic gain and thus the balancing of the subtraction must be done optically, through manipulation of the $\eta_{{\rm bs}}$ term. This design, to be known for the rest of this paper as the {\it current subtracting} design, is given in detail in \cite{VahlbruchPhD}. The current subtracting design can result in reduced classical noise due to the fact that all of the electronic components after the photodiodes are common and also mitigates another noise source, flicker noise, by design. Flicker noise is a noise source found in resistors that is proportional to the amount of current passing through the resistor, which results in a 1/f roll-up in noise towards low frequencies. The source of flicker noise is not well known, but it can be mitigated using wire wound or metal film type resistors. Flicker noise is described, and an investigation into the flicker noise present in many different resistors, is illustrated in the work of Seifert \cite{Seifert09.LN}.  Flat shot-noise achieved through both of these designs is shown in figure \ref{VolCurSN}, with flat shot noise in the variable gain design achieved via the use of low flicker noise metal electrode leadless face (MELF) type resistors.

Dark noise, the electronic noise of the detection system when no light is incident on the photodiodes, can also limit the measurement of shot-noise, and squeezing even more so, particularly at lower frequencies. The dark current of the photodiode and the current and voltage noise in the op-amps are the major contributors towards dark noise. Correct selection of readily available components ensures that the dark noise is not a problem. The level that shot-noise should be above the dark noise is dependent upon the amount of squeezing one wishes to measure. Typical detectors can be made and bought with the shot-noise approximately 10$\,$dB above the dark noise but more advanced designs can have the shot-noise around 20$\,$dB above the dark noise, in the tens of Hertz to tens of thousands of Hertz frequency band for optical power levels of around 1 milliwatt \cite{VahlbruchPhD}.

\section{Optics}\label{Optic}
\begin{figure}
  \begin{center}
  \includegraphics[width=6cm]{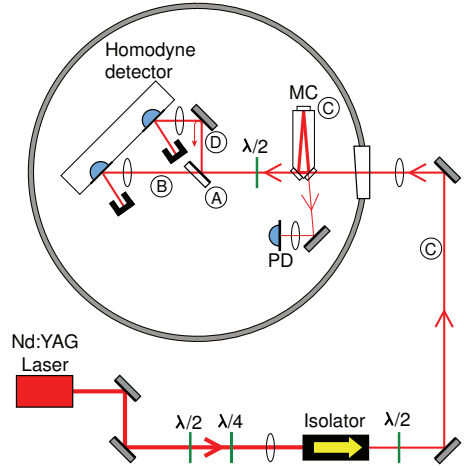}
  \end{center}
  \caption{The optical layout of the experiment. A high-speed photo-detector (PD) was used to lock the modecleaner (MC).
  The core components are enclosed in a steel anechoic chamber with an AR coated window. The circled labels refer to
  the locations of issues discussed in following sections (A) \ref{OptBal} - Optical Balancing (B) \ref{ScatterLoss} - Scattering Loss (C) \ref{BeamP} - Beam Pointing (D) \ref{ParasInt} - Parasitic  Interference.}
  \label{Layout}
\end{figure}

Many optical issues also need to be addressed in order to achieve flat shot-noise across the audio-frequency spectrum. These optical issues and their solutions were investigated on a simple tabletop experiment, shown in figure \ref{Layout}. The output of an ND:YAG 1064nm laser, after passing through a  Faraday isolator, was directed into a steel chamber through an anti-reflection (AR) coated window. Within the chamber, the light underwent spatial filtering with a triangular ring mode-cleaner cavity, or modecleaner, with a linewidth of approximately 4.7 MHz. The modecleaner was locked using the Pound-Drever-Hall technique \cite{Black01.AJP}. The light travels through a half-wave plate and is then incident on the beamsplitter which directs the light towards a current subtracting balanced homodyne detector \cite{VahlbruchPhD}. The half-wave plate is used to tune the beamsplitter power splitting ratio, $\eta_{{\rm bs}}$.

\subsection{Optical Balancing}\label{OptBal}

Balancing the optical splitting ratio of the beamsplitter is crucial to the performance of the balanced homodyne. A high Common Mode Rejection Ratio (CMRR), the measure of the ability of the balanced homodyne to reject signals common to both photodiodes, is necessary such that classical noise present on {\it both} fields entering the beamsplitter, the local oscillator and the signal field, is subtracted to a level well below the shot noise level by the balanced homodyne detector. This issue can be seen in (\ref{current1}) and (\ref{current2}). It is seen that with optimisation of the $\eta_{1,2}(x,y)$ and $\eta_{{\rm bs}}$ terms, the subtraction of these two photocurrents can completely remove all terms proportional to $\delta X_{a}^{+}$, which is the noise on the local oscillator field. This is necessary as we are only interested in measuring the signal field.

\begin{figure}
  \begin{center}
  \includegraphics[width=8cm]{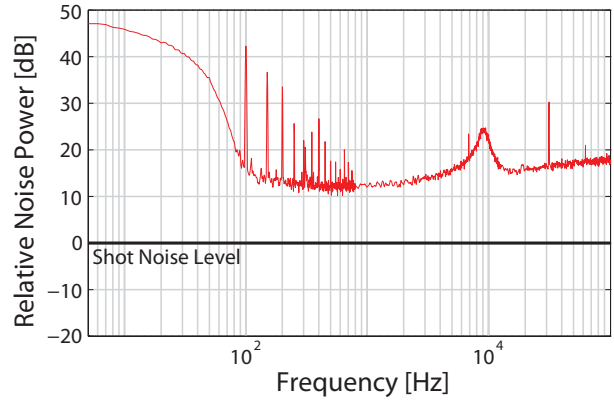}
  \end{center}
  \caption{Relative intensity noise of a 0.65$\,\mu$W field incident on one photodiode of the balanced homodyne detector. At
  low frequencies, local oscillator intensity noise is more than 40dB greater than shot-noise. The trace is pieced together from 3 FFT frequency ranges: 0-800$\,$Hz with 50 root mean square averages, 0-6.4$\,$kHz and 0-102.4$\,$kHz both with 100 averages. All traces taken with 800 FFT lines, resulting in resolution bandwidths of 1$\,$Hz, 8$\,$Hz and 128$\,$Hz respectively.}
  \label{IntNoise}
\end{figure}

The issue is compounded when the signal field itself carries noise which needs to be subtracted, such as in the coherent locking scheme \cite{Vahlbruch06.PRL}. Optimising the subtraction for the local oscillator arm, achieved by firstly tuning the angle of the beamsplitter and then through fine adjustment of either the variable gain or the polarisation of the fields, a stable CMRR of up to 80$\,$dB was consistently achieved with either balanced homodyne design. Figure \ref{IntNoise} shows that this level of subtraction should be adequate to remove all of the classical laser noise from the local oscillator across the entire spectrum. In this figure, beam pointing noise (See $\S$ \ref{BeamP}) accounts for much of the noise seen at low frequencies. A measurement showing the relative intensity noise of a similar laser with reduced beam pointing is shown by Nagano {\it et al.} \cite{Nagano98.LP}. With the CMRR optimised the local oscillator, a CMRR of 50$\,$dB was obtained for the signal arm. This is sufficient because the power in the signal arm is many orders of magnitude less than the local oscillator \cite{McKenzie04.PRL}.

\subsection{Scattering Loss}\label{ScatterLoss}

We define scattering loss as an event occurring in the beam path which causes the scattered light to exit the beam path. Scattering events before the beamsplitter will not affect the splitting ratio, but can couple noise in through changes in mode shape (This is further discussed in subsection \ref{BeamP}). However, an issue occurs if the loss originates in one of the balanced homodyne arms after the beamsplitter. This loss, $\eta_{l}$, is illustrated in figure \ref{Concept} and in equations (\ref{current1}) and (\ref{current2}). Dust particles, originating from many sources such clothing and skin, passing through the beam due to air currents or gravity will result in a non-zero loss term and momentarily reduce the magnitude of the photocurrent $i_{{\rm HD}1}$. This momentary change in the photocurrent occurs at frequencies characteristic of the speed of the particle and will unbalance the homodyne setup. This will result in a reduction of the CMRR, the temporary value of which may not be enough to cancel the classical intensity noise of the laser.

If the dust particle in the beam path reduces the power in one of the beams by one percent, the subtraction is reduced from $80\,$dB to $40\,$dB. Figure  \ref{IntNoise} shows that this reduction in CMRR is enough to couple classical intensity noise into the output of the balanced homodyne. In addition, intensity noise at frequencies characteristic of the time taken for the dust particle to traverse the beam, will couple directly to the output. Prior work has shown that settling of dust does in fact improve the noise statistics of the shot-noise \cite{Chua08.JPCS}.

Using the experimental setup shown in figure \ref{Layout}, it was determined that loss due to scattering could be a limiting factor by first setting the optical power on each photodiode to 5.7V, approximately 340 $\,\mu$W. With light on both photodiodes, the DC voltage was subtracted to less than 1 mV. An air puffer was then used to disturb the dust that had settled on various optics mounts. During this process, the subtracted output of the detector was monitored on an oscilloscope. With this configuration, substantial spikes in the subtracted output were observed when dust was excited with the puffer after the beamsplitter.

\begin{figure}
  \begin{center}
  \includegraphics[width=8cm]{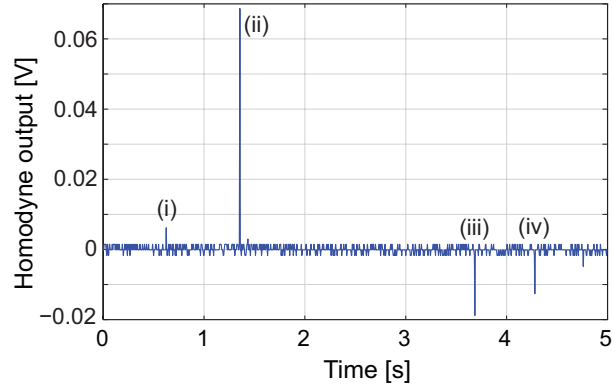}
  \end{center}
  \caption{The DC output of the balanced homodyne detector as a function of time. At time t=0 a puffer was used to excite dust on a lens mount in the beam path. The spikes in the output, (i)-(iv), are individual dust events affecting the balancing of the homodyne detector. Each photodiode has 5.7$\,$V, approximately 340$\,\mu$W, of incident light.}
  \label{DustvsTime}
\end{figure}

Figure \ref{DustvsTime} shows the voltage measured on the balanced homodyne detector over time after one such excitation. Four spikes were observed, the largest causing a 0.07$\,$V, or close to 1$\,$\%, change in the voltages between the two photodiodes. The typical voltage seen from these disturbances was approximately 0.01$\,$V. These spikes would occur over a few seconds following the excitation. 

We also excited dust activity before the beamsplitter to see what effect this has on the balanced homodyne. As expected, no spikes in the subtracted signal were observed because the dust-induced loss is now common to both arms. Isolation against scattering loss post homodyne beamsplitter is therefore crucial. The best way to stop dust is to ensure the cleanliness of the environment. Clean rooms, suitable clothing and enclosures will go a long way to reducing the amount of dust present in an experiment. However, there always will be some dust present and eliminating the air currents that carry the dust through the beam paths, such as with an enclosure, is the next step. The steel tank in our experiment both reduced the dust present and the air currents around the balanced homodyne detector. Once the tank was in place and sealed, the spikes in the DC output of the balanced homodyne detector were no longer seen.

\subsection{Beam Pointing and Photodiode Inhomogeneity}\label{BeamP}

A fundamental limitation to the photocurrent subtraction of the balanced homodyne detector is provided by the spatially non-uniform response of the two photodiodes in the detection scheme. This is represented by photodiode quantum efficiencies that are a function of position, $\eta_{1,2}(x,y)$. The surfaces of the photodiodes vary, and may include small ``dead'' regions on the semiconductor surface. Furthermore, investigations into the response of the surfaces has also shown that there are differences in the quantum efficiency of any photodiode across the photodiode surface \cite{KweePhD,SeifertPhD}. If the transverse profile or positioning of a beam on a photodiode surface changes, such as occurs due to beam pointing, then the photocurrent from the photodiode will also vary.

\begin{figure}
  \begin{center}
  \includegraphics[width=8cm]{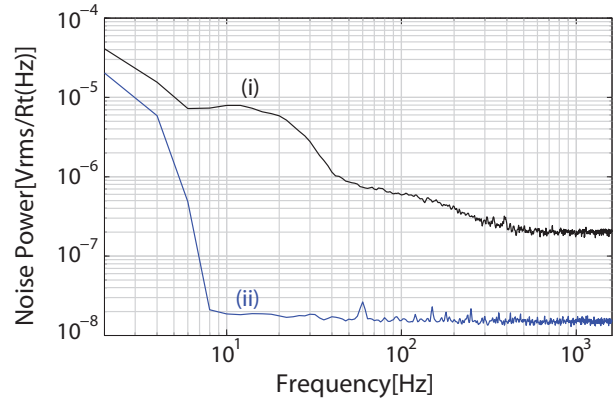}
  \end{center}
  \caption{Balanced homodyne output traces illustrating a beam pointing event. Trace (ii) is the regular shot-noise level with 1.3$\,$mW of local oscillator and trace (i) is the output of the balanced homodyne detector during a typical blow-out event. The experimental setup is that shown in figure \ref{Layout} albeit with the lid of the steel tank removed. Blow-outs such as that seen in trace (i) occurred every few seconds, with flat traces seen for most of the remaining time. All traces are measured with a span of 0-1.6$\,$kHz with 800 FFT lines, resulting in a resolution bandwidth of 2$\,$Hz. 100 root mean square averages were taken for both traces.}
  \label{Blowout}
\end{figure}

Beam pointing, also known as beam jitter, is a change in the trajectory of a beam and originates from forces acting on optical components, such as acoustic vibrations and air currents, or from changes in air density in the beam path resulting in varying refractive indices - the Schlieren effect \cite{Hecht}. Beam pointing noise is converted into photocurrent noise via the inhomogeneity of the two photodiode surfaces in the balanced homodyne detector \cite{Seifert96.OL}. The spectrum of this noise is determined by the specifics of the photodiode inhomogeneities and the speed of the beam translation. This noise is uncommon between the two photodiodes, and as such, it cannot be removed by photocurrent subtraction. The effect of beam pointing can be seen by disturbing the air near to the beam path, anywhere on the table. Many large blow-outs in the spectrum at frequencies up to about 1$\,$kHz can be seen in the output spectrum when this disturbance is introduced. One of these events is illustrated in figure \ref{Blowout}.

The coupling strength of beam pointing noise can be reduced by enlarging the beam spot size on the photodiode surface in order to effectively average out the photodiode inhomogeneities \cite{SeifertPhD}. However, the spot size is limited to a diameter where the quantum efficiency is not reduced by having portions of the beam undetected. Beam pointing itself can be reduced by designing the experiment to reduce pointing fluctuations by using short, shielded beam paths, and mechanically stable optics mounts.  To fully solve the issue a modecleaner was placed in the balanced homodyne detector setup \cite{Vahlbruch07.NJP}. Beam pointing can be thought of as the coupling of some fraction of the original beam power into different spatial modes. The cavity is designed such that these modes are non-degenerate, and as such the higher order modes do not resonate and are reflected from the cavity's input coupler. By this mechanism, pointing noise is converted into intensity noise. This intensity noise occurs {\it before} the beamsplitter and is now common to both arms of the homodyne detector, thus the large CMRR of the balanced homodyne will remove this intensity modulation. 

The modecleaner also filters changes in mode shape, which may occur due to dust passing through the beam. Changes in the mode shape also couple into intensity noise through the photodiode inhomogeneity. As with beam pointing, the modecleaner will convert mode shape fluctuations into intensity fluctuations which can then be subtracted.

Once the modecleaner was introduced, it was seen that air disturbances before the modecleaner did not result in low frequency blow-outs. The experiment is still sensitive to pointing noise which occurs downstream of the modecleaner. The steel tank was then used to stabilise air currents and reduce dust in this section of the experiment, solving the issues of beam pointing and varying mode shape.

\subsection{Parasitic Interference}\label{ParasInt}

Parasitic interferences, also called parasitic interferometers, were recognised as the limiting factor in the first measurements of flat shot-noise across the entire audio band \cite{Vahlbruch07.NJP}. Parasitic interference occurs when some light is scattered from the original beam path, through dust or surface imperfections, reflects off an object such as a mirror mount or another optical element, and re-enters the original beam. It is possible for the scattered light to exit the beam completely upon scattering and it is possible for the scattered light to reverse direction but travel within the original beam, sometimes called {\it small angle scatter}. Any change in the path length of the scattered field, typically caused by vibrations in the object that the scattered light is reflecting off, will modulate the intensity of the light. If the scattered light exits the beam path then careful placement of beam dumps will be enough to remove the resulting parasitic interference. However, if the scattered light does not exit the beam path, then it is more difficult to remove. The balanced homodyne detector is capable of measuring vacuum fluctuations, therefore small fractions of a single photon on average per bandwidth Hz are clearly visible.

\begin{figure}[h!]
  \begin{center}
  \includegraphics[width=9cm]{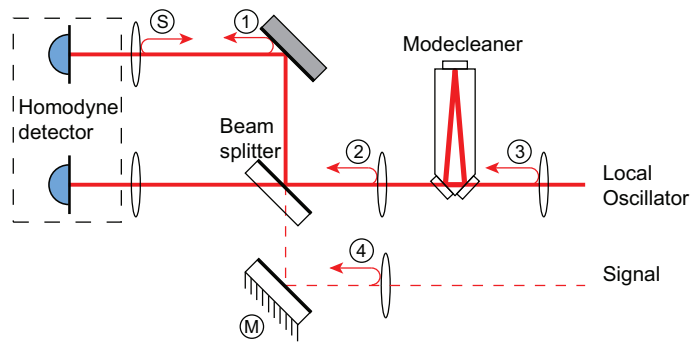}
  \end{center}
  \caption{Scatter in the balanced homodyne detector leading to parasitic interference, with a local oscillator power of 1.9$\,$mW. The backscattering occurs in one of the detector arms at location S and the scattered light travels along the beam path in the opposite direction. Forward scattering can then occur off a second surface, locations 1-4. The location of the forward scattering determines the effect on the balanced homodyne detector output, described in text.}
  \label{Scatter}
\end{figure}

Figure \ref{Scatter} shows the different locations at which scattering can occur. Here we have assumed that the scattered light does not leave the beam path but instead travels the same path in the opposite direction. Scattering is assumed to originate from either arm of the balanced homodyne detector and could occur at the photodiode or from one of the focussing lenses, illustrated in the figure \ref{Scatter} as location `S'. This is known as the {\it backscattered light}. When the backscattered field scatters for a second time, it is now known as the {\it forward scattered} field. This field travels in the same direction as the local oscillator field, resulting in a Mach-Zender-type interference between the forward scattered field and the local oscillator field. Figure \ref{Scatter} shows the possible locations where scattering can occur. We discuss the effect that parasitic interference, originating from different locations, will have on the shot-noise.

The most important thing to note is that interferometric effects that occur in the LO path between the beamsplitter and the laser source will result in intensity noise that is common to both arms of the balanced homodyne detector, which will be suppressed by the CMRR. This is the case for the forward scattering locations 2 and 3. Additionally, location 3 may be afforded some filtering effect by the modecleaner cavity, dependent upon how closely the mode of the scattered light matches the mode of the local oscillator field. The forward scattering seen at location 1 will set up a parasitic interference in only one arm of the balanced homodyne. The intensity noise on the optical field due to this parasitic interference is incident on only one photodiode, and hence cannot be subtracted. If the magnitude of this noise is large enough compared with the shot-noise, then it will be seen in the output spectrum. The final location is forward scattering from location 4, in the path where the signal field enters the balanced homodyne setup. Scattering from this location will have a substantial impact since the balanced homodyne system is designed to be maximally sensitive to any light that enters through the signal port. The squeezed vacuum that we wish to detect has no coherent amplitude and hence the scattered light will not interfere at the point of forward scattering. Instead, it will interfere at the beamsplitter with the local oscillator field and the intensity fluctuations will be anti-correlated in the two arms of the balanced homodyne, resulting in anti-correlated noise in each of the two photodiodes.

\subsection{Detecting the Presence of Parasitic Interference} \label{ParDet}

\begin{figure}[h!]
  \begin{center}
  \includegraphics[width=8cm]{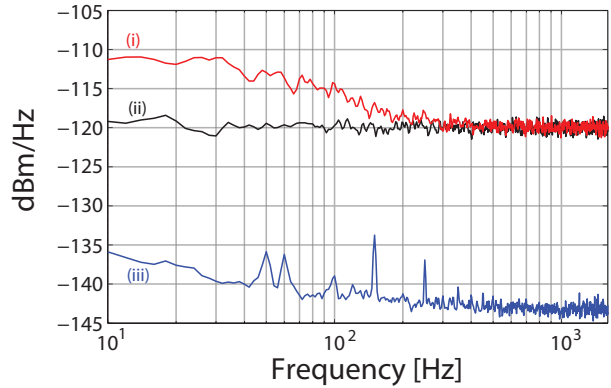}
  \end{center}
  \caption{Parasitic interference from the signal port of the balanced homodyne detector. Only the local oscillator field was present. Trace (i) is the homodyne output with the signal port open, trace (ii) is the balanced homodyne output with the beam dump in place and trace (iii) is the dark noise of the balanced homodyne detector.}
  \label{Fourthport}
\end{figure}

As a quick test to determine whether parasitic interference was present in our setup or nt, shot-noise spectra were taken with a beam dump in the signal port at location 4 in figure \ref{Scatter} and with the signal port open but with no laser fields present in this path, shown in figure \ref{Fourthport}. The beam dump was placed such that there were no optical components between it and the beamsplitter.

The difference between the shot-noise for the two cases shown in figure \ref{Fourthport} is strong evidence for the presence of parasitic interference. Flat shot-noise is seen when the beam dump is in place, at location 4 in figure \ref{Scatter}, but excess noise is seen when it is absent. A Faraday isolator with approximately 22dB of isolation was inserted in place of the beam dump but the excess noise did not noticeably change, indicating that greater isolation is required or perhaps that scatter off the Faraday isolator itself is enough to see these effects. The beam dump was moved up the signal path to try and determine if there was a single point where most of the forward scattering was occurring but it seemed to be a cumulative effect from several locations in this path.

The results from figure \ref{Fourthport} indicate that parasitic interference was likely present, but using a method known as opto-mechanical frequency shifting, or cyclic averaging, it is possible to determine with certainty whether a low frequency roll-up is due to the presence of scattered light and can provide information on the origin of the scattering \cite{Luck08.JOA}. The mirror M in figure \ref{Scatter} is actuated with a Piezo-Electric Transducer (PZT) that is driven with a triangle wave. Now consider what happens to light that is scattered from position S, backwards to position 4, where a scattering plate has been introduced, and then forwards to the balanced homodyne detection. The PZT is set up such that it sweeps the phase of the scattered light through an integer number of cycles, time-averaging the scattered light to zero. The signal due to the motion of the scatter sources now appears at the dither frequency of the PZT.

\begin{figure}[h!]
  \begin{center}
  \includegraphics[width=7cm]{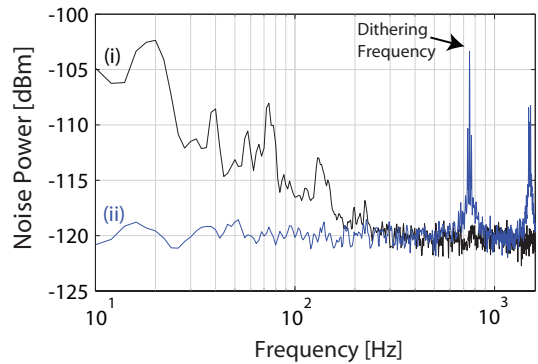}
  \end{center}
  \caption{The cyclic averaging technique applied to the balanced homodyne with a scatterer placed in the signal path. Trace (i) shows the measured shot-noise, with a local oscillator power of 1.9$\,$mW, when a scatterer was placed in location 4 in figure \ref{Scatter}. With this scatterer in place, the PZT placed between this scatterer and the detector's beamsplitter was dithered at a frequency of 750$\,$Hz. The amplitude of the dither was increased and adjusted until all of the low frequency noise was seen to shift up to the dither frequency and its harmonic as seen in trace (ii).}
  \label{ScatterDither}
\end{figure}

Figure \ref{ScatterDither} shows that the opto-mechanical frequency shifting-method has indeed confirmed the presence of scattered light. This is evident because the scattered light noise has been mixed up to the dithering frequency and the shot-noise is now flat below this frequency. This method has also provided us with additional information. The fact that the dithering results in flat shot-noise has provided information on the location of the scattering and it signifies that the scatter occurs in the beam path. If one were to move the position of the PZT in this beam path progressively further away from the balanced homodyne detector at some point it might be seen that the signal at the modulation frequency decreased. This would indicate that scattering has not undergone the phase shifts introduced by the PZT and hence scattering has occurred nearer to the balanced homodyne detector than the PZT.

Parasitic interference can be overcome in 3 ways; by reducing the amount of scattering by using components with minimal surface roughness, by dumping the scattered photons through careful placement of beam dumps, and by reducing the phase fluctuations in the scattered fields by reducing vibration in the experimental setup. The results presented in figure \ref{Squeeze} were achieved using super-polished optics and had an enclosure built around the entire experiment. It required irises in the beam path, careful dumping of the reflected fields from the photodiodes, dumping of reflected fields from AR coatings and transmitted fields through high-reflectivity (HR) mirrors and screens to isolate scattered light from different sections of the experiment.

\section{Measuring Squeezing}

\begin{figure}[h!]
  \begin{center}
  \includegraphics[width=9cm]{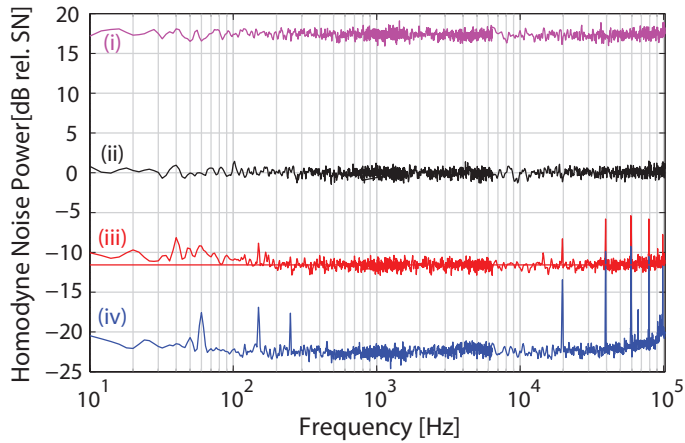}
  \end{center}
  \caption{Measured squeezing using quantum noise limited balanced homodyne detector showing the anti-squeezing (i); shot-noise (ii); squeezing (iii); and dark noise (iv). A red line at -11.6$\,$dB has been added to show the average of the squeezing from 1.6$\,$kHz to 6.4$\,$kHz. All traces are pieced together from 3 FFT windows: 0-1.6$\,$kHz, 0-6.4$\,$kHz, 0-102.4$\,$kHz with 800 FFT lines, resulting in resolution bandwidths of 2$\,$Hz, 8$\,$Hz, and 128$\,$Hz respectively. 100 RMS averages were taken for all traces. Dark noise has not been subtracted from the data.}
  \label{Squeeze}
\end{figure}

It has been shown that flat shot-noise can be achieved using the methods described in the previous sections. Using the doubly resonant, travelling wave optical parametric oscillator presented in a previous paper \cite{Chua10.OL} we measure the squeezing produced using the quantum noise limited balanced homodyne setup. The optical parametric oscillator is operated with 90$\,$mW of input power, 65$\,\%$ of threshold and the escape efficiency is measured to be 98.5$\pm$0.1$\,\%$. The expected propagation loss on the squeezed state is less than 0.7$\,\%$ and the balanced homodyne detector had a visibility of 99.4$\pm 0.2 \,\%$. The photodiodes are designed to have a quantum efficiency greater than 99$\,\%$ (Laser Components GmbH).  The results are shown in figure \ref{Squeeze}.

Figure \ref{Squeeze} shows that with a local oscillator power of approximately 1.9$\,$mW, the shot-noise of our detector is greater than 20$\,$dB above the electronic noise.  The anti-squeezing is 17.5$\,$dB above the shot noise and is flat. Below 200$\,$Hz, greater than 10$\,$dB of squeezing is seen and above this frequency 11.6$\,$dB of squeezing is observed. Correcting for dark noise, this squeezing level becomes 11.9$\,$dB. The squeezing and anti-squeezing levels observed correspond well to the predicted and measured efficiencies in the apparatus. The slight degradation in the measured squeezing below 200$\,$Hz indicates that there is still residual parasitic interference in our experimental setup.  The measurement thus shows that parasitic interference was suppressed to more than 10$\,$dB below shot noise.

\section{Conclusion}
Measuring flat shot-noise down to tens of hertz and below is a crucial requirement for measuring squeezing down to these same frequencies. Here we have presented an analysis of various issues which must be addressed in order to achieve the required performance. We have shown that, in accordance with previous results, to achieve reliable quantum noise limited balanced homodyne performance, the issues to consider are the balanced homodyne electronics design, dust, beam jitter, and scattered light leading to parasitic interference. A useful diagnostic tool, opto-mechanical frequency shifting, was used to both detect and aid in the mitigation of parasitic interference. With the quantum noise limited balanced homodyne setup, nearly flat levels of large-magnitude squeezing were seen across the entire frequency band of interest for advanced ground based interferometric gravitational wave detectors.

The authors would like to acknowledge financial support from the Australian Research Council the Deutsche Forschungsgemeinschaft, the Centre for Quantum Engineering and Space-Time Research (QUEST) and the IMPRS on Gravitational Wave Astronomy.

\section*{References}
\bibliographystyle{unsrt}
\bibliography{MasterRefMarch12}

\end{document}